\documentclass[12pt]{article}
\usepackage{amsmath,amsfonts, amssymb,graphicx,geometry,authblk,color,soul,comment,hyperref} 
\DeclareMathOperator{\argmin}{argmin}

\title{Effective behavior of heterogeneous media governed by strain gradient elasticity}
\author{Harkirat Singh, Mayank Raj, Kaushik Bhattacharya\footnote{Corresponding author.  Email: bhatta@caltech.edu}}
\affil{Division of Engineering and Applied Sciences,\\ California Institute of Technology, \\ Pasadena CA 91125}
\begin{document}
\maketitle
%%%%%%%%%%%%%%%%%%%%%%%%%%%%%%%%%%%%%%%%%%%%%%%
\begin{abstract}
Various mechanical phenomena depend on the length scale, and these have inspired a variety of nonlocal and higher gradient continuum theories.  Mechanistically, it is believed that the length scale dependence arises due to an interplay between the length scale of heterogeneities in the material, the length scale of the material being probed and the phenomenon under study.  In this paper, we seek to understand this interplay in a simple setting by studying the overall behavior of a one-dimensional periodic medium governed by strain gradient elasticity at the microstructural scale.  We find through numerical experiments that the overall behavior is not described by a strain gradient elasticity.  In other words, strain gradient theories are not invariant under averaging at this scale.  We also find that the overall behavior may be described by a kernel-based nonlocal elasticity theory, but the kernel is highly oscillatory with slow decay.  So we seek alternate characterization.  First, we limit our interest to a range of length scales, and show that the behavior is described  well by fractional strain gradient elasticity.  Consequently, one can obtain various scaling laws with exponent between zero (classical elasticity) and one (strain-gradient elasticity).  Second, we take a data-driven approach, and show that we can describe the overall behavior over a range of scales using a Fourier neural operator.
\end{abstract}

%%%%%%%%%%%%%%%%%%%%%%%%%%%%%%%%%%%%%%%%%%%%%%%
\section{Introduction}\label{sec:intro}

This paper is concerned with the overall behavior of heterogeneous materials where the underlying phenomenon depends on a length scale.  There are a number of observations of the dependence of mechanical properties on length scales, especially in plastic yield.  The seminal work is due to Hall \cite{hall1951deformation} and Petch \cite{petch1953cleavage} who showed that the yield strength increases with the inverse of the square-root of the grain size (smaller is stronger), and attributed this to dislocation pile-up at grain boundaries.  Fleck {\it et al.} \cite{fleck_1994} observed that the hardening in wires under torsion depends on the wire diameter while Nix and Gao \cite{nix1998indentation} observed length scale dependence in indentation; see also \cite{dunstan_2014} for a summary.  All of this has motivated a rich literature on gradient theories of plasticity \cite{aifantis_1987,fleck_1994,nix1998indentation,fleck_2001,gurtin_2005}.  Further scale effects have been observed in plasticity since the advent of nano-pillar compression \cite{greer_2011,mu_2014,mu_2016}.  A key conclusion of the newer observations is that the scaling law (the exponent in the strength or hardening vs.\ length scale) can vary widely depending on circumstance, and none of the gradient theories are able to predict these exactly.  This motivated Dahlberg and Ortiz \cite{dahlberg} to propose a theory of fractional strain gradient plasticity.  However, all of this modeling effort is macroscopic and phenomenological.  At the same time, the underlying mechanism for the length scale dependence is related to heterogeneities \cite{ashby_1970}.  Therefore, it would be interesting to understand the length scale dependence starting from a heterogeneous medium, and this is the motivation for the current work.

We consider the simple setting of one dimensional strain gradient elasticity in this paper.  The continuum theory of elasticity with length scale dependence takes one of two approaches.  The first is a nonlocal or ``integral'' approach (sometimes referred to as strongly nonlocal) where the stress at a point depends on the strain across the domain via decaying kernel functions \cite{eringen1972nonlocal,eringen_1972b,eringen1972linear}.  Such theories have been used to study (dispersive) wave propagation and defects like screw dislocations \cite{eringen1983differential}.  The second is the ``gradient'' approach (sometimes referred to as weakly nonlocal) wherein the stress is a function of the strain and its gradients at the same point \cite{koiter_1964,mindlin_1965,aifantis_1984,eringen1983differential}.   Such theories have been used to regularize singular solutions of classical elasticity, for example at crack tips \cite{vardoulakis1996gradient,zhou1999investigation}.  However, these theories are all postulated phenomenologically at the macroscopic scale.

There are also derivations of length scale dependent theories, especially strain gradient theories, starting from either discrete models using dispersion relations \cite{kunin_1966,krumhansl_1967}, or heterogeneous materials using ``higher order homogenization'' \cite{bakhvalov_1989,drugan_1996}.  A subtle issue that arises in the latter is that the truncation of the expansion at second order requires care, and a naive truncation can lead to a non-positive-definite second order modulus \cite{allaire_2016,smyshlyaev_2000,thbaut_2025}.  These derivations are formal, and valid in the asymptotic limit of zero ratio of microscopic (heterogeneity) to macroscopic length scale since they use Taylor expansion around the zero length scale ratio.

In this paper, we use numerical experiments to probe the effective behavior of a one-dimensional periodic medium governed by strain-gradient elasticity at the microscale.   Specifically, we have a medium with three scales: the material scale $\lambda$ due to the strain gradient, the microstructure scale $\ell$ or the periodicity of the underlying heterogeneous medium, and a macroscopic scale $L$ associated with the domain or the wavelength of loading.  We are interested in the situation where $\ell << L$.

M\"uller and Francfort \cite{muller_1994} studied the homogenization of such a medium under the asymptotic limit $\ell/L \to 0$ in multiple dimensions.  Adapted to one dimension, their result states that the homogenized behavior is classical elasticity (with no strain gradient behavior) with an effective elastic modulus $\bar{E}(\lambda/\ell)$ that depends on $\lambda/\ell$.  Further,
\begin{equation} \label{eq:fm}
\bar{E} (\lambda/\ell) \to \begin{cases}  \langle E^{-1} \rangle^{-1} \quad & \text{as } \lambda/\ell \to 0, \\ \langle E \rangle & \text{as } \lambda/\ell \to \infty \end{cases}
\end{equation}
where $\langle \cdot \rangle$ denotes the spatial average over the unit cell.  If $\lambda/\ell \to 0$, then there are no strain gradient effect, and the effective elastic modulus is the harmonic mean of the elastic modulus.  Conversely, if $\lambda/\ell \to \infty$, then the strain gradient dominates and we have a uniform strain field across the unit cell, and the effective elastic modulus is the arithmetic mean of the elastic modulus.  The effective elastic modulus increases monotonically from the harmonic to the arithmetic mean as $\lambda/\ell$ goes from zero to infinity.  In higher dimension, the effective elastic modulus is bounded by the effective modulus of classical elastic homogenization (no strain gradient) when $\lambda/\ell \to 0$, and volume average of the elastic modulus when $\lambda/\ell \to \infty$.   Explicit bounds for particular classes of microstructure have also been derived.  For example, Smyshlaev and Willis \cite{smyshlyaev_1994} extended the Hashin-Shtrikhman variational principle to the strain gradient setting, and obtained bounds on the effective shear modulus for a statistically isotropic two phase medium.

We study the correction to the asymptotic homogenized limit by conducting numerical studies of the situation where $\ell << L$, but the ratio $\ell/L$ is finite (away from the asymptotic homogenized limit).  In practice, we have a finite ratio $\ell/L$, and this finite ratio can be fit to any power of $\ell/L$.  So one is never sure at which power one should truncate the expansion.  Therefore, it is of practical interest to understand the behavior near a finite, but small, ratio $\ell/L$.  In particular, we would like to understand if such a situation is described by a strain gradient theory, or a more general nonlocal theory.

We find that the overall behavior is not described by strain gradient elasticity.  In other words, strain gradient theories are not invariant under change of scale.  Instead, the overall behavior may be described by a kernel-based nonlocal elasticity theory.  Unfortunately, we find that the kernel is extremely oscillatory, and therefore difficult to use in the solution of problems.  Therefore, we seek two alternate characterizations of the effective behavior.

The first is explicit, but focusses on a limited range of length scales.   We show that the effective behavior is described by a fractional strain gradient in any range of length scales.  We study a simple boundary value problem, and show that the overall behavior  displays a power-dependance in this range of length scales.  Crucially, the scaling exponent can vary between zero and one depending on the reference length scale of the range.  The exponent is zero (classical elasticity) when the reference length scale is small and one (strain gradient theory) when the reference length scale is large.  We are unaware of any other microscopic derivation of a fractional strain gradient theory.

The second is a data-driven representation in the form of a neural approximation.  In recent years, various researchers have used neural networks as constitutive relations (see for example \cite{flaschel_2025} and the references there).  Deep neural networks are a highly expressive class of functions from one finite dimensional vector space to another.   Indeed, they can approximate any continuous function on a compact set.  Thus, it provides a rich framework to develop constitutive relations in settings like elasticity.  However, in a nonlocal setting, the constitutive relation is a map from the strain field to the stress field, or  an operator from one infinite dimensional space to another.  One may seek a finite dimensional map from a discretized strain field to a discretized stress field, but, unfortunately,  this neural network approximation would be specific to that discretization, and would lose accuracy in any other discretization.  Neural operators \cite{kovachki2023} are a generalization of neural networks to an infinite dimensional setting, and can be used independent of discretization.  We show that we can approximate the nonlocal effective behavior using a Fourier neural operator \cite{li_2021,kovachki2023,bhattacharya_learning_2024}.  

The paper is organized as follows.  We introduce our one-dimensional strain gradient theory in Section \ref{sec:sge}, and describe methods we use to solve problems in Section \ref{sec:methods}.  We gain insight into strain gradient elasticity by studying a finite hanging bar (a bar fixed at one end, free at the other and subjected to uniform body force) in Section \ref{sec:hang}.  The heart of the paper is Section \ref{sec:periodic_bar} where we study an infinite bar subjected to periodic forcing.  Section \ref{sec:learn} describes operator learning.  We conclude in Section \ref{sec:conc}.

%%%%%%%%%%%%%%%%%%%%%%%%%%%%%%%%%%%%%%%%%%%%%%%
\section{Strain-gradient elasticity}\label{sec:sge}

We consider a bar of length $\tilde L$ and uniform cross-section $\tilde A$, possibly made of a periodic microstructure with unit cell length $\tilde \ell$.  The elastic modulus is $\tilde E$, and the strain gradient modulus is $\tilde \kappa$.  We have an imposed body force per unit length $\tilde f$.  The total energy in the system undergoing displacement $\tilde u$ is
\begin{eqnarray}
\tilde{\mathcal E}(\tilde u) = \frac{1}{2} \int^{\tilde L}_{0} \tilde A \left( \tilde E \tilde u_{\tilde x}^2 + \tilde \kappa \tilde u_{\tilde x \tilde x}^2 - \tilde f \tilde u \right)d \tilde x + \widetilde{\mathcal B},     
\end{eqnarray}
where $\widetilde{\mathcal B}$ depends on the boundary condition.  Note that $\tilde E, \tilde \kappa$ may be heterogeneous (periodic with period $\tilde \ell$).    It is convenient to non-dimensionalize using a reference length $\tilde L_0$ for length and a reference modulus $\tilde E_0$ for energy density ($\tilde E_0 \tilde L_0^2$ for $\tilde \kappa$ and $\tilde E_0/ \tilde L_0$ for $\tilde f$).  We denote the non-dimensional quantities by the name letters without the tilde.  The energy now is 
\begin{eqnarray}
{\mathcal E}(u) = \frac{1}{2} \int^{L}_{0} A \left( E u_{x}^2 + \kappa u_{xx}^2 - f u\right)dx + {\mathcal B}
= \frac{1}{2} \int^{L}_{0}  A \left(E (u_{x}^2 + \lambda^2 u_{xx}^2) - f u \right)dx + {\mathcal B}
\end{eqnarray}
where $\lambda = \sqrt{\kappa/E}$ is the (non-dimensional) material length.  

Note have three (non-dimensional) lengths:  the material length $\lambda$, the unit cell period $\ell$ and the macroscopic length $L$.  We can make one of these lengths 1 by a choice of $\tilde L_0$, but choose not to do so till later.

The resulting equilibrium equation is
\begin{eqnarray} \label{eq:ge}
\left(E u_{x} - (\lambda^2E u_{xx})_{x}\right)_{x} + f =0,     \label{eq:eq1}
\end{eqnarray}
and the jump conditions are
\begin{eqnarray} \label{eq:jump}
[[ u ]]=0, \,\, [[ u_x ]]=0, \,\, [[ E u_x - (\kappa u_{xx})_{x} ]]=0,\,\, [[ \kappa u_{xx}]]=0.
\end{eqnarray}
%and the natural boundary conditions are
%\begin{eqnarray} \label{eq:nbc}
%(\kappa u_{xx})_{x} (0)=0   \,\,\, \text{and} \,\,\, (\kappa u_{xx})_{x} (L)=0.
%\end{eqnarray}
as shown in Appendix \ref{app:eqj}.   Note that we have two stresses, the Cauchy stress $E u_x - (\kappa u_{xx})_{x}$ and a higher order stress $ \kappa u_{xx}$.  We need to add appropriate boundary conditions.  

We study two problems in the following sections, a finite length bar subject to uniform body force in Section \ref{sec:hang} and an infinite bar subject to periodic loading in Section \ref{sec:periodic_bar}.  But, first, we describe the numerical methods to solve the governing equations.

\section{Methods}\label{sec:methods}

%%%%%%%%%%%%%%%%%%%%%%%%%%%%%%%%%%%%%%%%%%%%%%%
\subsection{Transfer matrix method}

We seek to solve (\ref{eq:eq1}) on $(0,L)$ subject to boundary conditions at $x=0,L$ for a bar whose the moduli is piecewise constant,
\begin{equation}
E(x) = E_i, \  \kappa(x) = \kappa_i \quad x \in (x_{i-1},x_i), \quad \quad  i=1,\dots ,I
\end{equation}
where $x_0=0 < x_1 < \dots < x_I=L$,   using a transfer matrix method \cite{haskell_1953,knopoff_1964}.  In each segment $(x_{i-1},x_i)$, it is easy to verify that the homogeneous solution to (\ref{eq:eq1}) is given by
\begin{equation}
u(x) = c_1 \exp \left(\frac{x}{\lambda_i} \right) + c_2 \exp \left(-\frac{x}{\lambda_i} \right) + c_3 x + c_4 .
\end{equation}
Set $\varphi_i = (u,u_x,Eu_x - (\kappa u_{xx})_x, \kappa u_{xx})^T$ in the $i^\text{th}$ segment.  Using the solution above, we may write
\begin{equation} \label{eq:itrans0}
\varphi_i(x_{i-1}) = \Phi_i (x_{i-1}) C + \varphi^p_i (x_{i-1}), \quad 
\varphi_i(x_{i}) = \Phi_i (x_{i}) C + \varphi^p_i (x_{i}), \quad 
\end{equation}
where $\Phi$ is the matrix given in Appendix \ref{app:tm} and $C = (c_1,c_2,c_3,c_4)^T$.  $\varphi^p$ is obtained from the particular solution depending on $f$, and expressions for constant and sinusoidal $f$ are given in Appendix \ref{app:tm}.  Eliminating $C$, we find
\begin{equation} \label{eq:itrans}
\left(\varphi_i(x_i) - \varphi_i^p(x_i) \right) = T_i \left(\varphi_i(x_{i-1}) - \varphi_i^p(x_{i-1}) \right)
\end{equation}
where $T_i  = \Phi_i (x_i) \Phi_i(x_{i-1})^{-1} $ is the transfer matrix in the $i^\text{th}$ segment.

Now, the jump conditions (\ref{eq:jump}) implies that $\varphi$ is continuous, i.e., $\varphi_i(x_{i-1}) = \varphi_{i-1}(x_{i-1})$.  We can use this to infer that
\begin{equation}
\begin{aligned}
\left(\varphi_i(x_i) - \varphi_i^p(x_i) \right) &=
T_i \left(\varphi_{i-1}(x_{i-1}) - \varphi_{i-1}^p(x_{i-1}) \right) - T_i [[ \varphi^p (x_{i-1}) ]]  \\
&= T_iT_{i-1} \left(\varphi_{i-1}(x_{i-2}) - \varphi_{i-1}^p(x_{i-1}) \right) - T_i [[ \varphi^p (x_{i-1}) ]] \label{eq:trans}\\
&= T_iT_{i-1}T_{i-2} \left(\varphi_{i-2}(x_{i-3}) - \varphi_{i-2}^p(x_{i-3}) \right)  \\
& \quad \quad  - T_iT_{i-1} [[\varphi^p(x_{i-2})]] - T_i [[\varphi^p(x_{i-1})]]  \\
&= \left(\Pi_{\alpha=1}^i T_\alpha \right)  \left(\varphi_{1}(0) - \varphi_{1}^p(x_{0}) \right)
- \sum_{\beta=1}^i \left(\Pi_{\alpha=\beta}^i T_\alpha \right)  [[\varphi^p(x_{\alpha-1})]].  
 \end{aligned}
\end{equation}
Above, we apply continuity of $\varphi$ at $x_{i-1}$ to obtain the first equality,  the analog of (\ref{eq:itrans}) for $(i-1)$ to obtain the second and so forth.  Setting $i=I$, the relationship (\ref{eq:trans}) enables us to relate $\varphi(L)$ to the solution  at $\varphi(0)$.  Since four of these eight quantities are known from the boundary conditions, we may regard this as a set of linear equations for the remaining unknowns.

We can also use the transfer matrix method in the periodic setting by enforcing $\varphi(L) = \varphi(0)$ from periodicity and solving (\ref{eq:trans}) for $\varphi(0)$.

%%%%%%%%%%%%%%%%%%%%%%%%%%%%%%%%%%%%%%%%%%%%%%%
\subsection{Fast Fourier transform method}

We consider a periodic domain and a periodic forcing $f$.  We follow Moulinec and Suquet \cite{moulinec_1994,moulinec_1998} and others (see \cite{moulinec_2018} for a recent review) to look for a periodic solution to the governing equation (\ref{eq:eq1}) using a Lippmann-Schwinger type iterative method,
\begin{equation}
E_0 u^{n+1}_{xx} - \kappa_0 u^{n+1}_{xxxx} = -f - \left( (E-E_0) u^n_{x}\right)_{x} + \left((\kappa-\kappa_0) u^n_{xx}\right)_{xx} =: {\mathcal F}^n,   
\end{equation}
where $E_0>0, \kappa_0>0$ are uniform constants.  Taking the Fourier transform,
\begin{equation}
(- k^2 E_0 - k^4 \kappa_0)\hat{u}^{n+1} (k) = \widehat{ \mathcal F}^n (k)  \quad \implies \quad \hat{u}^{n+1}(k)  = \frac{\widehat{ \mathcal F}^n (k) }{(k^2 E_0+ k^4 \kappa_0)}
\end{equation}
where $k$ is the variable in the Fourier domain.

Given $u^n$ and its derivatives, we compute ${\mathcal F}^n$ in real space and then use fast Fourier transform (FFT) to compute $\widehat{ \mathcal F}^n$.  This enables the calculation of $\hat{u}^{n+1}$; an inverse FFT provides $u^{n+1}$.  We iterate in this manner until we converge according to the criteria,
\begin{align}
r_1 = \mid \mid u^{n+1} - u^{n} \mid \mid \le r^{\rm tol}_1, \quad
r_2 = \mid \mid u_{xx}^{n+1} - u_{xx}^{n} \mid \mid \le r^{\rm tol}_2
\label{convergence}
\end{align} 
for given tolerances $ r^{\rm tol}_1,  r^{\rm tol}_2$.

%%%%%%%%%%%%%%%%%%%%%%%%%%%%%%%%%%%%%%%%%%%%%%%
\subsection{Bloch-Floquet theory}

In a periodic setting where the forcing has a larger period than the unit cell, we can invoke the Bloch-Floquet theorem that a $2\pi/k$ periodic function $u$ may be represented as 
\begin{equation}
u(x) = v(x) e^{ikx},
\label{eq:bloch-floquet}
\end{equation}
where $v(x)$ is $\ell-$periodic (e.g., \cite{ashcroft_1976}).  We can use this in both the transfer matrix and fast Fourier transform method to solve the problem on the unit cell.

%%%%%%%%%%%%%%%%%%%%%%%%%%%%%%%%%%%%%%%%%%%%%%%
\section{Hanging bar}\label{sec:hang}

We begin with a simple boundary value problem in a bar of finite length $\tilde L$.  It is convenient to normalize all lengths with  $\tilde L$ ($\tilde L_0 = \tilde L$ and $L=1$) in this section.

A  bar is suspended from one end, free at the other and subject to a uniform body force.  So, we set the displacement to be zero on one end, the stress to be zero at the other, and the higher order stress to be zero on both ends:
\begin{eqnarray} \label{eq:hang_bc}
u(0) = u_{xx}(0)=0, \quad  u_{xx}(1) = (E u_x - (\kappa u_{xx})_x) (1)=0, \quad f=f^* ,
\end{eqnarray}
where $f^*$ is a constant.

%%%%%%%%%%%%%%%%%%%%%%%%%%%%%%%%%%%%%%%%%%%%%%%
\subsection{Homogeneous bar}

\begin{figure}
\centering
\includegraphics[width=5in]{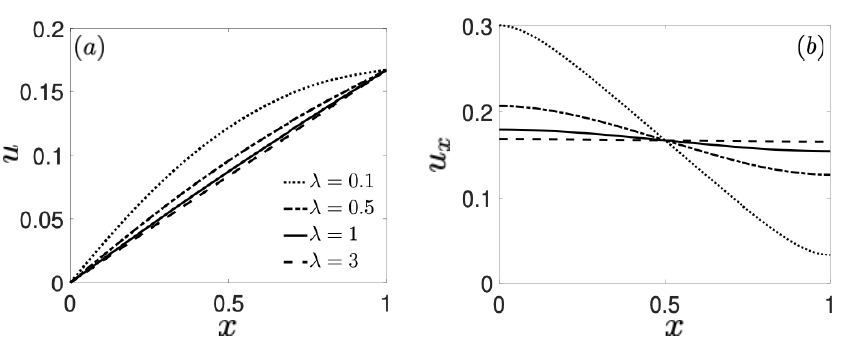}
\caption{Displacement (a) and strain (b) in a homogeneous hanging bar for various material length scales $\lambda$.  All lengths are normalized with the length of bar $L$. }
\label{fig:homo}
\end{figure}

We solve (\ref{eq:eq1}) subject to the boundary condition and body force (\ref{eq:hang_bc}).  It is easy to verify that the solution is given by
\begin{eqnarray}
u(x) = \frac{f^* \lambda^2}{E} \left( \frac{e^{x/\lambda}}{1  + e^{1/\lambda}}  + \frac{e^{- x/\lambda}}{1  + e^{- 1/\lambda}} -1 \right) +  \frac{f^*}{2 E}(2 x - x^2) .
\end{eqnarray}
Figure \ref{fig:homo} displays the displacement and strain for various values of material length  $\lambda = \sqrt{\kappa/E}$.  Note that the solution depends on the material length.   Still, the end displacement, $u(1) = f^* /2 E$, is independent of the higher order modulus $\kappa$, and thus the material length.  This is a consequence of the boundary conditions (\ref{eq:hang_bc}), and specifically the fact that  the higher order stress is zero at both ends.  Other boundary conditions would have given a dependance on the material length (and indeed it is common to choose boundary conditions to obtain the desired scaling in gradient theories).  However, our goal is to understand the interplay between material length and heterogeneity, and this boundary condition eliminates the effect of the boundary condition.

%%%%%%%%%%%%%%%%%%%%%%%%%%%%%%%%%%%%%%%%%%%%%%%
\subsection{Heterogeneous bar}

We now consider the same problem but with periodic moduli given by 
\begin{eqnarray} \label{eq:per}
E(x) = \begin{cases} E_1 & 0 \le \left\{\frac{x}{\ell}\right\} < 1/2 \\
E_2 & 1/2 \le \left\{\frac{x}{\ell}\right\} < 1 \\
\end{cases} \ , \quad
\kappa(x) = \begin{cases} \kappa_1 & 0 \le \left\{\frac{x}{\ell}\right\} < 1/2 \\
\kappa_2 & 1/2 \le \left\{\frac{x}{\ell}\right\} < 1 \\
\end{cases} \ ,
\end{eqnarray}
where $\{ y \}$ denotes the fractional part of $y$ (the difference between $y$ and the largest integer smaller than $y$), $\ell$ is the period of the microstructure (so the bar has $N=1/\ell$ unit cells since we have normalized all lengths by the length of the bar in this section).

We solve it using the transfer matrix method.   For future use, we denote the quantity 
\begin{equation} \label{eq:bare}
\bar{E} = \frac{f^*}{2 u(1)}
\end{equation}
the {\it effective elastic modulus} motivated by the relation between the modulus and end displacement in the case of the homogenous bar.

\begin{figure}[t]
\centering
\includegraphics[width=6in]{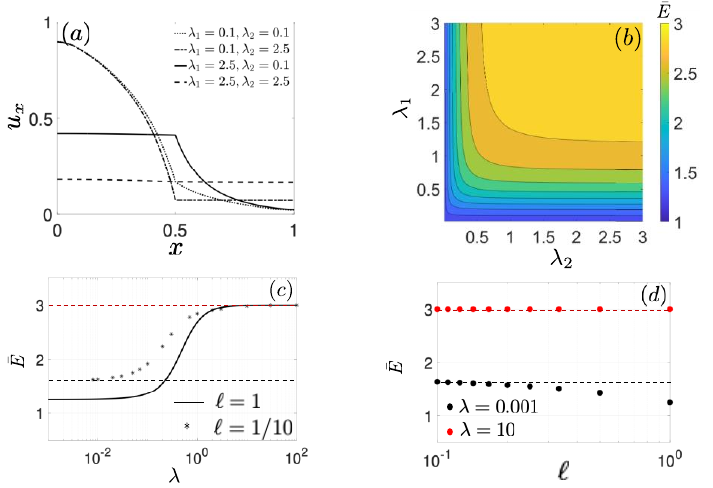}
\caption{Strain and effective elastic modulus of a heterogeneous bar with $E_1=1$ and $E_2=5$. (a) Strain in a two-piece bar ($\ell=1$) for various combinations of material lengths $\lambda_1, \lambda_2$. (b) Effective elastic modulus of a two-piece bar ($\ell=1$) for various combinations of material lengths.  (c) Effective elastic modulus for $\ell=1, 1/10$ as a function of the material length when $\lambda_1=\lambda_2=\lambda$.  (d)  Effective elastic modulus as a function of $\ell$ for large and small material lengths ($\lambda_1=\lambda_2$).  The upper red dashed line is the arithmetic mean 3 and lower black dashed line is the harmonic mean 5/3 of the elastic moduli.  All lengths and normalized so that the length of the bar is 1.}
\label{fig:hetero}
\end{figure}

We start with the case of a two-piece bar where $\ell=1$ ($N=1$) and fix $E_1=1$, $E_2=5$.  Figure \ref{fig:hetero}(a) shows the  strain for various combinations of material lengths $\lambda_1$ and $\lambda_2$.  We see the effect of the interface, and its dependance on the material lengths.  The effective modulus $\bar{E}$ (cf.\ (\ref{eq:bare})) is shown in Figure \ref{fig:hetero}(b).  In contrast to the case of the homogeneous bar, the effective modulus of the two-piece bar depends on the material scales due to the presence of the boundary layers near the interface.  When both the material lengths are large, these boundary layers dominate the bar and so the effective elastic modulus approaches arithmetic mean of the two elastic moduli ($\bar{E} = \langle E \rangle = 3$).  In contrast, when either or both material lengths are small, the boundary layer at the interface is small and we approach the elastic solution ($\bar{E} = \langle E^{-1} \rangle^{-1} = 5/3$).   In fact, when $\lambda_1=\lambda_2=\lambda$, we can show that 
\begin{equation}
\bar{E} (\lambda) = \frac{15}{4} \left(3 - 6 \lambda^2 + 6\lambda^2 \, {\rm sech}\left(\frac{1}{2 \lambda} \right)  - 2 \lambda \tanh\left(\frac{1}{2 \lambda} \right) \right)^{-1}.
\end{equation}
This is shown in the Figure \ref{fig:hetero}(c) as the solid line.

We then consider a periodic bar with $N$ periods ($\ell=1/N$) in the case of equal material lengths, $\lambda_1=\lambda_2= \lambda$.  Figure \ref{fig:hetero}(c) shows the effect of $\lambda$ for the cases $\ell =1, 1/10$ ($N=1,10$).  We see that for the case $\ell =1/10$, the effective elastic modulus interpolates between the harmonic and arithmetic means as $\lambda$ goes from zero to $10^2$.  This is as we expect from the analysis of M\"uller and Francfort \cite{muller_1994} in the limit $\ell \to 0$ described in the introduction (specifically, (\ref{eq:fm})).   

More importantly, we see from Figure \ref{fig:hetero}(c) that the effective elastic modulus for the case $\ell=1$ is different from that in the case $\ell=1/10$.  This means that the effective elastic modulus can depend on $\ell$ for a fixed $\lambda$.  We explore this further in Figure \ref{fig:hetero}(d) that shows how the effective elastic modulus changes with $\ell$ in the bar for two fixed values of material scale $\lambda$.  Note that the limit of small $\ell$ is consistent with the results of the M\"uller and Francfort \cite{muller_1994} in (\ref{eq:fm}).    When the material lengths are large, the boundary layer dominates and the effective elastic modulus remains close to the arithmetic mean with no significant dependence on $\ell$.  In contrast, when the material lengths are small, we see that the effective modulus decreases with $\ell$.  To understand this, recall that we approach the elastic solution as $\lambda$ is small.  Therefore in a two segment bar with $\ell=1$, we have a larger strain in the first segment since  $E_1<E_2$ even though it has  a smaller stress under uniform body force.  This results in a smaller effective elastic modulus.  This effect diminishes with decreasing $\ell$ (increasing number of unit cells), and the effective elastic modulus approaches the  harmonic mean $\bar{E} = \langle E^{-1} \rangle^{-1}$ corresponding to a purely elastic bar.

In summary, this simple  boundary value problem shows that the overall response of a heterogeneous material governed microscopically by strain gradient elasticity has a nontrivial dependence on the ratio between the material heterogeneity and the macroscopic scale.  We explore this systematically in the next section.

%%%%%%%%%%%%%%%%%%%%%%%%%%%%%%%%%%%%%%%%%%%%%%%
\section{Infinite bar subjected to periodic forcing} \label{sec:periodic_bar}

We now characterize the macroscopic behavior of a heterogeneous material with microscopic strain gradients.  We do so by considering an infinite bar, and subject it to a periodic body force.  It is convenient in this setting to normalize all lengths by the length of the unit cell (so $\tilde L_0 = \tilde \ell$ and $\ell = 1$ in this section).  We specifically consider the body force
\begin{equation} \label{eq:bf}
f = \sin kx \quad \text{where} \quad k = \frac{2\pi}{L}
\end{equation}
and $k$ is the wave number and $L$ is the wavelength of loading.  We assume that $L$ is rational ($L=a/b$ for $a,b$ integers) so that both the microstructure and forcing are periodic with period $a=bL$.  So we look for solutions with period $a$, but can use the Bloch-Floquet theory to reduce the problem to the unit domain $(0,1)$.

Note that this wavelength $L$ is the macroscopic length in this setting.  We are interested in understanding the situation when $k < 2\pi$ ($\tilde \ell/ \tilde L < 1$)  finite.

%%%%%%%%%%%%%%%%%%%%%%%%%%%%%%%%%%%%%%%%%%%%%%%
\subsection{Homogeneous bar}

First, we consider an infinite homogeneous bar subject to the body force (\ref{eq:bf}).  It is easy to see (using the Fourier transform, for example) that the solution to the governing equation (\ref{eq:ge}) is given by
\begin{align}
    u(x) = \frac{\sin(k x)}{E {k}^{2} + \lambda^2E{k}^{4}}
    \label{eq:sol_pbc_homo}
\end{align}
up to a constant.  The solution scales as $k^{-2}$ corresponding to classical elasticity when $k \to 0$ (the forcing is slowly varying and so the strain is uniform on the scale that the strain gradient acts), and  $k^{-4}$ corresponding to strain gradient theory when $k$ is large (the forcing and hence the strain is rapidly varying  on the scale that the strain gradient acts).

%%%%%%%%%%%%%%%%%%%%%%%%%%%%%%%%%%%%%%%%%%%%%%%
\subsection{Hetergenenous bar}

\begin{figure}[t]
\centering
\includegraphics[width=6in]{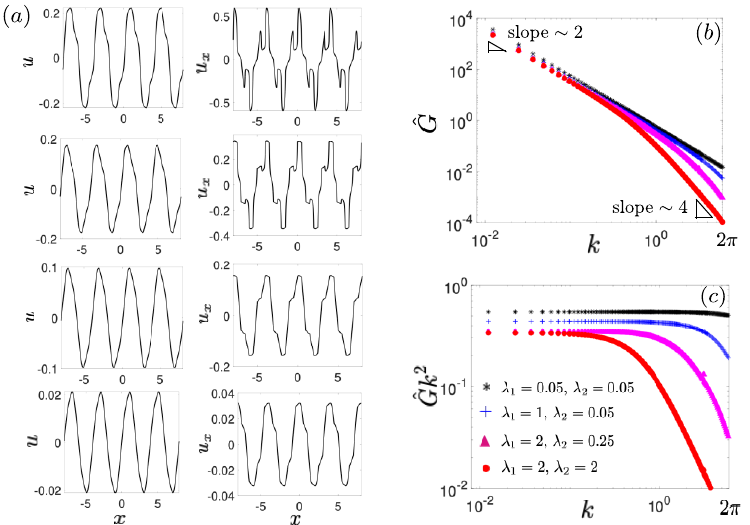}
\caption{Response of a periodic heterogeneous bar with  $E_1=1, E_2=5$ subjected to sinusoidal loading (a) Displacement and strain over four loading wavelengths with $k=1$ ($L=2\pi$), (b) Amplification $\hat{G}$ and (c) $\hat{G}k^2$ for various material length scales. All parts have the same combination and order of material lengths $\lambda$ as stated in the legend of (c),  and all lengths and normalized so that the unit cell length $\ell=1$.  We fix $E_1=1, E_2=5$}
\label{fig:hetper}
\end{figure}

We now consider the same problem, but with the periodic moduli distribution given by (\ref{eq:per}).  As stated above, we normalize all lengths with the period $\ell$, and hence are interested in wave numbers $k<2 \pi$ so that the loading wave-length is larger than the unit cell.  We fix $E_1=1, E_2=5$.  We solve the problem (\ref{eq:eq1}) using both the transfer matrix method and fast Fourier transform method, and verify that they result in the same solution.

Figure \ref{fig:hetper}(a) shows the displacement and strain when $k=1$ or $L=2\pi$ for four combinations of material length.  We observe that the dominant response is at the wave number $k=1$, though there are finer oscillations due to the heterogeneity.   We focus on the dominant response.  Indeed, since (\ref{eq:eq1}) is linear and elliptic, we expect that we can write the solution as
\begin{equation}
u(x) = (G*f)(x) = \int G(x-y)f(y) dy
\end{equation}
for a Green's function $G$.  Taking the Fourier transform,
\begin{equation}\label{eq:uhat}
\hat{u}(k) = \hat{G}(k) \hat{f}(k).
\end{equation}
The invariance $f(x) \to f(-x)$ implies $u(x) \to u(-x))$ due to frame-indifference, and this in turn implies that $\hat{G}$ is real or $G$ is symmetric about the origin.
%Therefore, 
%\begin{equation}
%\hat{G} = \int f(x) \cos{2\pi kx} dx .
%\end{equation}
%\hl{(Check normalization.)}

\paragraph{Overall response}
Figure (\ref{fig:hetper})(b) shows the response $\hat{G}$ as a function of the wave number $k$ at scales larger than the unit cell for four combinations of material scales.  Note that this figure is plotted on a logarithmic scale.  In all the cases, the large length scale limit ($k \to 0$) has a slope of $2$.  It follows,
\begin{equation} \label{eq:kto0}
- \lim_{k \to 0} \frac{d \log \hat{G}}{d \log k} = 2 \iff \hat{G} \approx \frac{1}{\bar{E} k^2} \text{ for small }k.
\end{equation}
This is emphasized in Figure (\ref{fig:hetper})(c) that shows $\hat{G}k^2$ as a function of $k$.  We conclude that the long wavelength limiting behavior is elasticity for some effective elastic modulus $\bar{E}$.  This is consistent with the analysis of  M\"uller and Francfort \cite{muller_1994}, and our results in the hanging bar.  We will study $\bar{E}$ in detail later.

Figure (\ref{fig:hetper})(b) also shows that the slope approaches $4$ as $k \to 2\pi$ when the material lengths $\lambda$ are large.  This is because the gradient term dominates the response in this regime and the strain is almost uniform in the unit cell.

\paragraph{Effective theory}
\begin{figure}
\centering
\includegraphics[width=6.5in]{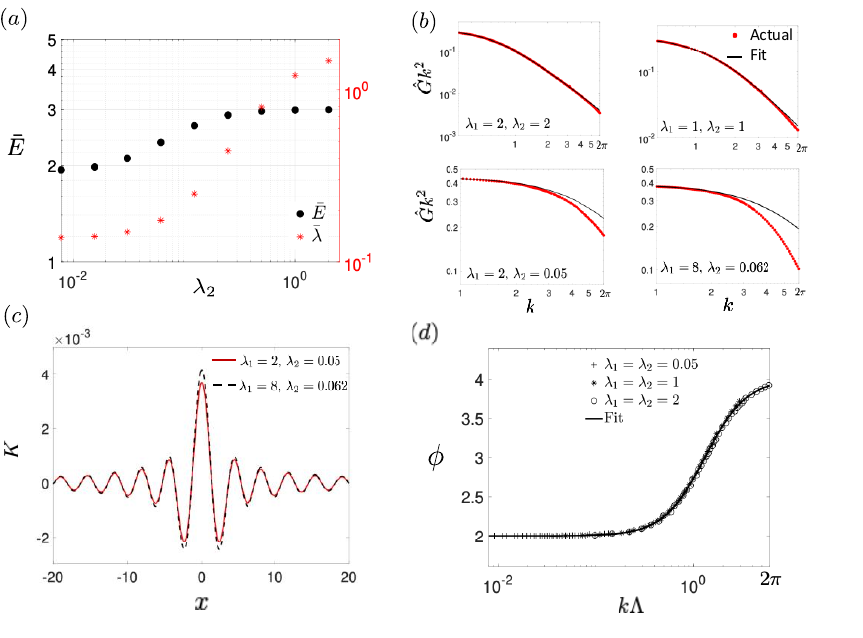}
\caption{The effective response of a heterogeneous bar.  (a) The best fit effective elastic modulus $\bar{E}$ and effective material length $\bar{\lambda}$ as a function of the material length $\lambda_2$ with $\lambda_1=2$.  (b) Comparison between the best effective strain gradient elasticity and actual response. (c) The kernels of nonlocal elasticity that describe the effective behavior a few combination of material lengths  and (d) Empirical relationship between the exponent $\phi$ (cf. (\ref{eq:phi}) and scaled wavenumber $ k\Lambda$.  $E_1=1, E_2=5$.}
\label{fig:eff}
\end{figure}
The asymptotic result of Drugan and Willis  \cite{drugan_1996} states that the effective response follows an effective strain gradient theory, or 
\begin{equation}
\hat{G} \approx \frac{1}{\bar{E} {k}^{2} + \bar{\lambda}^2 \bar{E} {k}^{4}}
\label{eq:esge}
\end{equation}
with an effective elastic modulus $\bar{E}$ and effective material length $\bar{\lambda}$.
We fit $\bar{E}$ and $\bar{\lambda}$ at $k=0$:
\begin{equation} \label{eq:ek}
\bar{E} =  \lim_{k\to0} \ (\hat{G}(k) k^2)^{-1}, \quad \bar{\lambda}^2 = \lim_{k \to 0} \ \frac{(\hat{G}k^2 \bar{E})^{-1}-1}{k^2}
\end{equation}
($\bar{E}$ is also the from the intercept of Figure (\ref{fig:hetper})(c)).
The results are shown in Figure \ref{fig:eff}(a) as a function of $\lambda_2$ when $\lambda_1=2$.  The elastic modulus $\bar{E}$ varies from the harmonic mean to the arithmetic mean consistent with the results of M\"uller and Francfort \cite{muller_1994}, and our results in the hanging bar.  The effective material length $\bar{\lambda}$ increases with $\lambda_2$: still, note that $\bar{\lambda} \ne 2$ even when $\lambda_1 = \lambda_2 = 2$.  This is because our elastic modulus is not uniform.  In other words, heterogeneous elastic modulus affects the effective material length.

We then compare the effective strain gradient elasticity theory according to (\ref{eq:esge}, \ref{eq:ek}) with the computed effective response in Figure \ref{fig:eff}(b).  We see reasonable agreement when the two material lengths are the same, but see a deviation when the material lengths are different.  Thus, {\it the effective behavior of a heterogeneous strain gradient elastic medium is not described by a strain gradient elastic theory}.  In other words, the interaction between the different material lengths give rise to interactions that are not described by a strain gradient theory.
We note that this behavior is not a pathology of the fact that we have discontinuous material parameters.  We show examples of smooth material heterogeneity in Appendix \ref{app:smooth}.

\paragraph{Nonlocal elasticity}
The general representation is given by a nonlocal elasticity theory \cite{eringen1972linear} where the stress at a point depends on an integral of the strain
\begin{equation}
\sigma (x) = \bar{E} \varepsilon (x) + (K*(K*\varepsilon))(x) = \bar{E} \varepsilon (x) + \int K(x-z)\int K(z-y) \varepsilon(y) dy \ dz
\end{equation}
for a suitable kernel $K$.  The corresponding equilibrium equation is
\begin{equation} \label{eq:nonloc}
\left( \bar{E} u_x + K*(K*u_x) \right)_x + f = 0.
\end{equation}
Taking the Fourier transform of this equation and recalling the definition of the Green's function, we can calculate the Fourier transform of the kernel in terms of the Fourier transform of the Green's function
\begin{equation}
\hat{G} = \frac{1}{k^2(\bar{E} + \hat{K}^2)} \iff \hat{K} =   \left(\frac{1}{\hat{G}k^2} -  \bar{E} \right)^{1/2}.
\end{equation}
We can now use our results to obtain the Fourier transform of the kernel, and then the kernel.  The result is shown in Figure \ref{fig:eff}(c) for a few combinations of material length scales.  Note that the kernel is extremely oscillatory.  This is quite different from the relatively smooth Gaussian-like kernels that have been used in the literature.  Further, the oscillatory nature makes the use of such kernels very difficult in practice.  This motivates us to look for an alternate representations.  The first, presented in the next paragraph, is explicit but for a narrow range of $k$ ($\ell/L$).  The second, presented in the next section, is based on a neural operator and is valid for all $k<1$.

\paragraph{Fractional strain derivative elasticity}

We are often interested in a narrow range of length scales.  Indeed many experiments on length scale effects are limited to factors of two to five  \cite{greer_2011,mu_2014,mu_2016}.  In this situation, it is natural to look for the response in this limited range of length scales.  We do so by formally linearizing the Fourier transform of the Green's function $\hat{G}$ and wave number $k$ in the logarithmic scale around a specific wave number $k_0$.
\begin{equation} \label{eq:exp}
\log \hat{G}(k) \approx \log \hat{G}(k_0) - \phi(k_0) (\log k - \log k_0)
\end{equation}
where
\begin{equation} \label{eq:phi}
\phi (k_0) := - \frac{d \log \hat{G}}{d \log k} (k_0).
\end{equation}
We find empirically that $\phi \in (2,4)$, and it fits the relation
\begin{equation}
    \phi (k_0) = 2 + 2 \left( \left(\frac{R_{\rm tr}}{\Lambda k_0}\right)^m + 1 \right)^{-1}
    \label{eq:hyp_tan}
\end{equation}
where $\Lambda$ and $m$ are fitting parameters, and  $R_\text{tr} = 0.212$.  This relationship is shown in Figure \ref{fig:eff}(d), and the fitting parameters for some cases are listed in Table \ref{tab:par}.   In the case when $\lambda_1=\lambda_2$, we find that $\Lambda = \lambda_1=\lambda_2, m = 2$.

\begin{table}[t]
\centering
\begin{tabular}{|c|c|c|c|}
\hline
$\lambda_1$&$\lambda_2$&$\Lambda$&$m$\\
\hline
$\lambda$ & $\lambda$ & $\lambda$ & 2\\
\hline
1 & 0.05 & 0.048 & 3.27\\
\hline
8 &0.0625 &0.0541& 3.82\\
\hline
\end{tabular}
\caption{Fitting parameters in the universal relationship between the exponent $\phi$ and wavenumber $k$.  $E_1=1, E_2 =5$.}
\label{tab:par}
\end{table}

Returning to (\ref{eq:exp}), we can rewrite it as
\begin{equation} \label{eq:gfrac}
\hat{G}(k) \approx \hat{G}(k_0) \left(\frac{k}{k_0} \right)^{-\phi(k_0)} = \frac{C}{2M \Lambda^{\phi-2} k^\phi}.
\end{equation}
for some effective modulus $M$ (with units of energy per volume) and some effective material length $\Lambda$. $C$ is a constant that depends only on $\phi$.  Recalling that the strain $\varepsilon$ in this one-dimensional setting is the derivative of the displacement, and combining (\ref{eq:gfrac}) with (\ref{eq:uhat}), we obtain
\begin{equation}
\hat{\varepsilon} = ik \hat{G} \hat{f} \implies \frac{2M\Lambda^{\phi-2} |k|^\phi }{C} \hat{\varepsilon} = ik \hat{f},
\end{equation}
or
\begin{equation}
ik \left( \frac{2M\Lambda^{\phi-2} }{C}  |k|^{\phi-2} \hat{\varepsilon} \right) +  \hat{f} = 0.
\end{equation}
Taking the inverse Fourier transform, we recognize this equation in real space as the equilibrium equation
\begin{equation} \label{eq:shat}
\sigma_x + f = 0 \quad \text{where} \quad \hat \sigma = \frac{2M\Lambda^{\phi-2} }{C}  |k|^{\phi-2} \hat{\varepsilon}.
\end{equation}
We have obtained the nonlocal effective stress-strain relation in Fourier space.

We can interpret this relation in real space using the theory of fractional calculus \cite{adams_2003,tartar_2007} (see Di Nezza {\it et al} \cite{dinezza_2012} for an easily accessible introduction).  We define the fractional Laplacian of order $\alpha \in (0,1)$ as 
\begin{equation}
(- \Delta)^\alpha u  := C \int_0^L \left(\frac{ u'(x) - u'(y) }{|x-y|^{1+2\alpha}} \right) dy.
\end{equation}
where the constant $C$ depends only on $\alpha$.  The Fourier transform gives (see \cite{dinezza_2012}),
\begin{equation}
\widehat{(- \Delta)^\alpha u} = |k|^{2\alpha} \hat u .
\end{equation}
Comparing this to (\ref{eq:shat}), we obtain 
\begin{equation}
\sigma = \frac{2 M \Lambda^{2\alpha}}{C} (- \Delta)^\alpha \varepsilon, \quad \text{where} \quad \alpha = \frac{\phi-2}{2} \in (0,1).
\end{equation}
Thus, the stress is linear in the $\alpha^\text{th}$ fractional Laplacian.

Equivalently, as shown in Appendix \ref{app:fd}, we can pose the equilibrium problem 
as one of minimizing the total energy where the total energy is defined using the $\alpha$-fractional derivative through the Gagliardo 2-semi-norm,:
\begin{equation} \label{eq:varfrac}
\min_u  
{\mathcal E}(u), \quad {\mathcal E}(u) = \int^{L}_{0}A  \left(\int^{L}_{0} \frac{1}{2} M \Lambda^{2\alpha} \frac{( u'(x) - u'(y) )^2}{|x-y|^{1+2\alpha}} dy - f(x)u(x) \right)dx + \mathcal{B}
\end{equation}
where $\mathcal B$ depends on the boundary condition.

We now use this variational formulation to study a simple problem where a bar is subjected to a dead load to understand the resulting scaling law.  The approach follows a calculation by Dahlberg and Ortiz \cite{dahlberg}.  We take $f=0$, and 
\begin{equation}
\mathcal{B} = A \sigma (u(L)-u(0)) = A \sigma \int_0^L u'(x) dx
\end{equation}
corresponding to a dead traction $\sigma$ at the ends of the bar.  Substituting this in (\ref{eq:varfrac}), we have
\begin{equation}
\begin{aligned}
\min_u \mathcal{E} (u) &= \min_\varepsilon  \int^{L}_{0}A  \left(\int^{L}_{0} \frac{1}{2} M \Lambda^{2\alpha} \frac{( \varepsilon(x) - \varepsilon(y) )^2}{|x-y|^{1+2\alpha}} dy - \sigma \varepsilon (x) \right)dx\\
& =  AL \min_\varepsilon \int^{1}_{0} \left(  \frac{1}{2} M  \Lambda^{2\alpha} L^ {-2\alpha} \int^{1}_{0} \frac{1}{2} \frac{( \varepsilon(\bar{x}) - \varepsilon(\bar{y}) )^2}{|\bar{x}-\bar{y}|^{1+2\alpha}} d \bar{y} - \sigma \varepsilon \right)d\bar{x} .
\end{aligned}
\end{equation}
Now, given any $\varepsilon(x)$, let $\bar{\varepsilon} = (\int_0^1 \varepsilon(\bar x) d\bar x) $, and $\tilde \varepsilon = \varepsilon/\bar{\varepsilon}$, so $\varepsilon = \bar\varepsilon \tilde \varepsilon$ with $\int_0^1 \tilde \varepsilon d\bar x = 1$.   Note that $\bar{\varepsilon} =  (\int_0^L \varepsilon(x) dx )/L $ is the average strain in the bar. Substituting this back, we find
\begin{equation} \label{eq:outer}
\begin{aligned}
\min_u \mathcal{E} (u) &=  \min_{\bar{\varepsilon}} \ AL \left( \frac{1}{2}M  \Lambda^{2\alpha} L^ {-2\alpha} \bar{\varepsilon}^2 D  - \sigma \bar \varepsilon \right) ,
\end{aligned}
\end{equation}
where
\begin{equation}
D = \min_{ \left\{\tilde \varepsilon:  \int_0^1 \tilde \varepsilon d\bar x = 1 \right\} } \ \int^{1}_{0}  \frac{( \varepsilon(\bar{x}) - \varepsilon(\bar{y}) )^2}{|\bar{x}-\bar{y}|^{1+2\alpha}} d \bar{x} 
\end{equation}
is independent of $L$.  We conclude from (\ref{eq:outer}) that the average strain,
\begin{equation}
\bar \varepsilon = \frac{\sigma}{MD} \left(\frac{\Lambda}{L} \right)^{2 \alpha},
\end{equation}
depends on the size with exponent $2 \alpha$.  When $\alpha = 0$ ($\phi=2$) there is no dependance on length scale and we are in classical elasticity.  This occurs when $k \to 0$ (cf. Figure \ref{fig:eff}(d)) in agreement with the result of M\"uller and Francfort \cite{muller_1994}.   Similarly, when $\alpha =1$ ($\phi = 4$) we have quadratic dependance corresponding to purely strain gradient elasticity.    In general, $\alpha \in (0,1)$ and we have an intermediate scaling law.  Further, recalling that $\alpha$ is related to the reference wavelength $k_0$, we can observe a  range of  scaling laws in a material depending on the range of macroscopic length scales we examine.

In summary, we conclude that {\it the effective behavior of a heterogeneous strain gradient elastic medium can be approximated by a fractional strain gradient elasticity in a limited range of length scales.   Further, the effective behavior displays a power-dependance on scale in this range of length scales.  Crucially, the  scaling exponent can vary between zero and one depending on the reference length scale.  It is zero (classical elasticity) when the reference length scale is small and one (strain gradient theory) when the reference length scale is large}. 

%%%%%%%%%%%%%%%%%%%%%%%%%%%%%%%%%%%%%%%%%%%%%%%
\section{Operator learning} \label{sec:learn}

We conclude this work by seeking a data-driven characterization of the effective behavior of a periodic medium that is valid over a whole range of length scales using a neural approximation.  Deep neural networks are an extremely versatile class of functions.  Indeed, according to the universal approximation theorems, we can show that given any continuous function, and any compact set, we can find a deep neural network that can approximate the continuous function up to any desired accuracy on the compact set.  Further, they have a structure that makes it possible to very efficiently fit these functions to given data.  Finally, the availability of open source resources makes it very easy to adopt and implement.  Therefore, they have been found to be extremely useful in a variety of fields, including increasingly in mechanics.  There are a number of excellent introductory notes and textbooks to deep neural networks including Strang \cite{strang_2019}.  Given the versatility, they have been used to model constitutive behavior of materials, especially in elasticity (see for example \cite{flaschel_2025} and the references there).

Note that deep neural networks are functions that map an input in a finite dimensional space to an output in a finite dimensional space.  So, they are ideally suited for a local setting like classical elasticity where we need a map from deformation gradient to stress.  However, we found that our effective behavior is nonlocal, and therefore the constitutive relation is an operator that maps an infinite dimensional strain field to an infinite dimensional stress field.  Thus, one cannot use a deep neural network that is limited to finite dimensions.  One may consider discretizing the strain and strain fields to obtain a finite dimensional function, but then the resulting neural network approximation is valid only for that discretization.  It leads to uncontrolled error at any other discretization.  This is unsatisfactory for a constitutive relation.

This motivates a neural operator, that is a generalization of neural networks to operators \cite{bhattacharya_2021,li_2021,kovachki2023}.  The key idea is to embed information of the underlying function spaces, so that any discrete data is regarded as a discretization of a function rather than simply discrete data.
We work with a particular neural operator, the Fourier neural operator (FNO), and show that it is able to learn the overall response of the material.  We specialize to the case of equal material scales, $\lambda_1=\lambda_2=\lambda$, fix $E_1=1, E_2=5$, and seek to learn the operator $\{f, \lambda\} \to u$.  We continue to normalize lengths so that $\ell = 1$

%%%%%%%%%%%%%%%%%%%%%%%%%%%%%%%%%%%%%%%%%%%%%%%
\subsection{Learning from data and Fourier neural operator}

We seek to approximate an unknown or expensive operator $\Phi: X \to Y$, where $X$ and $Y$ are Hilbert spaces, from data  $\{i_a, o_a = \Phi(i_a) \}_{a=1}^A \subset X \oplus Y$  obtained by $A$ evaluations of the unknown operator $\Phi$.  We do so by choosing a parametrized approximation $o = \Psi(i, P)$ that is highly expressive or versatile, and then estimate the parameter $P$ to find the best fit to the data by minimizing a loss function that depends on the difference between the prediction and data.  

Typically, the data is split into three parts, training ($A_1$ pairs), validation ($A_2$ pairs) and test ($A_3$ pairs).  The training dataset is used in the definition of the loss function. So,
\begin{equation}
P = \argmin \mathcal{L}, \quad \mathcal{L} = \sum_{a=1}^{A_1} || o_a - \Psi(i_a,P) || ,
\end{equation}
for an appropriate norm $|| \cdot ||$.  It is common to take the $L2$ norm and we do so as well.  The optimization is performed by stochastic gradient descent over a number of iterations or epochs.  The validation dataset is used to decide the end of training.  Finally, the test dataset that is completely unused in the training (choice of $P$) is used to test the accuracy of the resulting approximation.

We choose a {\it Fourier neural operator} (FNO) \cite{li_2021,kovachki2023} as our parametrized approximation.  We define this operator $i \to o$ through the relation,
\begin{equation} \label{eq:fno}
\begin{aligned}
&o = \left(  {\mathcal Q} \circ g(v_{H-1}) \circ \cdots  \circ g(v_{h-1}) \circ \cdots  \circ g(v_0)  \circ {\mathcal P} \right) i,\\
&\quad \quad \text{where} \quad v_h = g(W_n v_{h-1} + {\mathcal F}^{-1}(R_{n}){\mathcal F} v_{h-1} + b_n), \quad h = 1, \dots, H, 
\end{aligned}
\end{equation}
that is a composition of an input layer $\mathcal P$, $H$  hidden layers $g(v_{h-1}), h= 1, \dots H $, and an output layer $\mathcal Q$ in the language of neural networks.  The input $i$ is discrete  in $d_i$ dimensions and $\mathcal P$ is a lifting  from $d_i$ to $d_0$ dimensions.  We interpret the input as the discretization of a function, for example through the introduction of a basis (or more precisely by regarding $\mathcal P$ as Nemytskii operator).  At any  $h = 1, \dots, H$, $v_{h-1}$ is a vector in $d_{h-1}$ dimension, ${\mathcal F}$ and ${\mathcal F}^{-1}$ denote Fourier and  inverse Fourier transforms and $W_h, R_h$ are $d_{h-1} \times d_h$ matrices (weights) and $b_h$ is a $d_h$ dimensional vector (bias).  So, $u_h =W_h v_{h-1} + {\mathcal F}^{-1}(R_{h}){\mathcal F} v_{h-1} + b_h$ is a $d_n$ dimensional vector.  $g$ is the activation function, a continuous monotone function to be chosen, and this is applied to each component of $u_h$.  Finally $\mathcal Q$ is a projection  from $d_n$ dimensions to $d_o$ dimensions.  The output is also interpreted as the discretization of a function, for example through the introduction of a basis (or more precisely by regarding $\mathcal Q$ as Nemytskii operator).  The unknown parameters are $P = \{W_n, R_n, b_n\}$.  $H$ is called the depth of the network and the largest dimension $d_h$ is called the width or number of channels.

The form (\ref{eq:fno}) closely follows that of neural networks with two crucial differences.  First, the input and output are interpreted as discretization of a function by regarding $\mathcal P, \mathcal Q$ as Nemytskii operators.  Second, we have a Fourier transform within each layer.   The FNO is highly versatile, and there are universal approximation theorems not only when $X,Y$ are Hilbert spaces but also Sobolev spaces.  Further,  the efficacy has been demonstrated in Navier-Stokes \cite{kovachki2023}, in elliptic problems \cite{bhattacharya_learning_2024} and other applications.

In this work, we use the Gaussian error linear unit (GeLU) activation function,
\begin{equation}
g(x) = \frac{x}{2} \left(1+ \text{erf}\frac{x}{\sqrt{2}} \right)
\end{equation}
(erf is the Gaussian error function), the adaptive moment estimation (Adam) optimizer for training \cite{Adam}, and implement it in the package PyTorch \cite{Pytorch}.

%%%%%%%%%%%%%%%%%%%%%%%%%%%%%%%%%%%%%%%%%%%%%%%
\subsection{Learning the linear response}

\begin{figure}[t]
\centering
\includegraphics[width=6in]{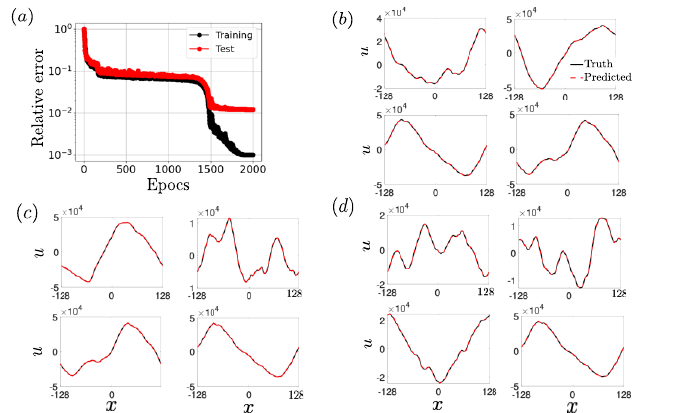}
\caption{Learning the linear map.  (a) Training and test error.  (b) Comparison of the displacement field for four instances of test data.  (c) Comparison of the displacement field for four instances of test data generated at a higher resolution than that used for training. (d) Comparison of the displacement field for four instances of test data generated at a lower resolution than that used for training.  $E_1=1, E_2=5$.}
\label{fig:fno_lin}
\end{figure}

We begin with the linear map $f \to u$ ($i=f, o = u$).  We use three hidden Fourier layers ($H=3$) with 256 Fourier modes and 64 channels ($d_h=64$) in each layer.  We generate 600 samples $(\{f_a,u_a\}_{a=1}^{A=600}$  by choosing a random forcing $f_a$ and computing the resulting $u_a$ at a resolution of 1024.  The training and test errors over 2000 epochs (iterations of the training optimizer) are shown in Figure \ref{fig:fno_lin}(a).   Figure \ref{fig:fno_lin}(b) compares the computed (truth) and predicted displacements for four randomly chosen test instances (these were not used in the training or validation).   We see that the trained FNO reproduces the results for a previously unseen input accurately.  As noted above, an important property of FNOs is that it learns the operator and not the discretized map even though it is trained on discretized data.  This is shown in Figures \ref{fig:fno_lin}(c) and (d) where the already trained FNO is evaluated against data generated at a resolution of 4096 and 512 respectively.  We have also verified linearity and the ability of the FNO to reproduce $\hat{G}(k)$.

%%%%%%%%%%%%%%%%%%%%%%%%%%%%%%%%%%%%%%%%%%%%%%%
\subsection{Learning the material and response}

\begin{figure}[t]
\centering
\includegraphics[width=6in]{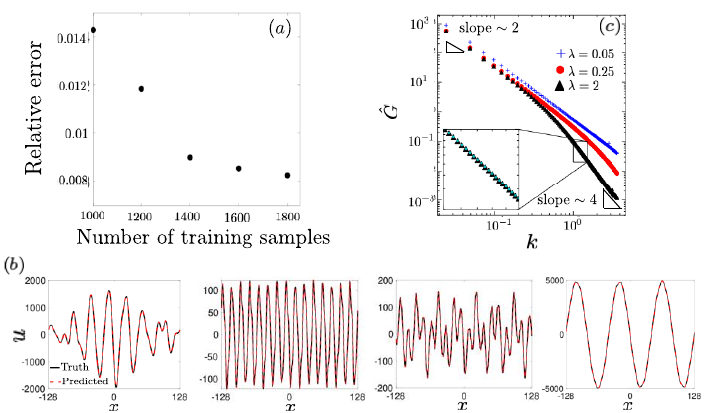}
\caption{Learning the material and response.  (a) Test error.  (b) Comparison of the displacement field for four instances of test data.  (c) Comparison of the Fourier transform of the Green's function.  $E_1=1, E_2=5$. Green circles in the inset are the actual simulations.}
\label{fig:fno}
\end{figure}

We now turn to the nonlinear map  $\{f,\lambda\} \to u$  ($i=\{f,\lambda\}, o = u$).  The results are shown in Figure \ref{fig:fno}.  We create a dataset of 2500 samples with 500 each at $\lambda = {2, 1, 0.5, 0.25, 0.05}$.  We pick 2000 samples at random for training and validation, and retain the remaining $500$ samples for testing.  We use 2000 epochs for training.  Figure \ref{fig:fno}(a) shows the training error while Figure \ref{fig:fno}(b) compares four test samples.  Figure \ref{fig:fno}(c) compares the computed spectrum of the Green's function $\hat{G}(k)$ (truth) with that predicted by the FNO. We see excellent match showing that the FNO is able to learn the effect of the material length scale. 

%\hl{It would be good to redo this section by generating data for random $\lambda_i, E_i$ (and possible volume fractions).}

In summary, we conclude that {\it the effective behavior of a heterogeneous strain gradient elastic medium can be learnt using a Fourier neural operator}.

\section{Conclusion} \label{sec:conc}

Motivated by the observation that nonlocal theories are a result of heterogeneity, we have studied a simple problem of a one-dimensional heterogeneous medium governed by strain gradient elasticity at the microstructural scale.  We have characterized  the overall behavior on scales that are large compared to the microstructural scale. We show that the overall behavior is not described by strain gradient elasticity.  In other words, strain gradient theories are not invariant under change of scale.  We also show that the overall behavior may be described by a kernel-based nonlocal elasticity theory, and locally approximated by a fractional strain gradient elasticity.  Consequently, one can obtain various scaling laws with exponent between zero (classical elasticity) and one (strain-gradient elasticity).  Finally, we have shown that the overall behavior may be learned using a Fourier neural operator.   We plan to build on this to study plasticity and other phenomena in future work.

%%%%%%%%%%%%%%%%%%%%%%%%%%%%%%%%%%%%%%%%%%%%%%%
\section*{Acknowledgements}
We gratefully acknowledge the financial support of the Office of Naval Research through MURI award N00014-23-1-2654.

%%%%%%%%%%%%%%%%%%%%%%%%%%%%%%%%%%%%%%%%%%%%%%%

%%%%%%%%%%%%%%%%%%%%%%%%%%%%%%%%%%%%%%%%%%%%%%%
\appendix
\section*{Appendix}
\renewcommand{\thefigure}{A\arabic{figure}}
\setcounter{figure}{0}
\renewcommand{\theequation}{A\arabic{equation}}
\setcounter{equation}{1}

%%%%%%%%%%%%%%%%%%%%%%%%%%%%%%%%%%%%%%%%%%%%%%%
\section{Equilibrium, jump and boundary conditions} \label{app:eqj}
Assuming the elastic bar of length $L$ made of two materials with a single interface at $l$, with zero displacement at one end, and an traction $\sigma$ at the other,  the energy can be written as 
\begin{eqnarray} \label{eq:ei}
{\mathcal E}(u) = \frac{1}{2} \left(\int^{l}_{0} \left( E u_{x}^2 + \kappa u_{xx}^2 - 2 f u\right)dx + \int^{L}_{l} \left( E u_{x}^2 + \kappa u_{xx}^2 - 2 f u\right)dx \right)dx     
- \sigma u(L) .
\end{eqnarray}
The above expression for the energy is well defined only when $u$ and $u_x$ are continuous  which implies that $[[ u ]]_{x=l} =[[ u_x ]]_{x=l}=0$. Minimizing the energy with respect to  $u$ implies $ \delta_u E = \frac{\partial}{\partial \epsilon} E(u+ \epsilon v)\big|_{\epsilon=0}$ 
 $\forall\,\,v$ s.t $v(0)=0$. Applying this to the energy integral (\ref{eq:ei}), and integrating by parts, we conclude:
\begin{equation}
\begin{aligned}
&- \int^{L}_{0}\left(\left( ( E u_x -(\kappa u_{xx})_{x}  \right)_{x} + f\right)v dx \\ 
& \quad \quad - [[ \left(E u_x - (\kappa u_{xx})_{x} \right)v ]]_{x=l}  - [[ \left(\kappa u_{xx} \right)v_x ]]_{x=l} \\
& \quad \quad + ( ( Eu_x - (\kappa u_xx)_x - \sigma) v)_{x=L} + (\kappa u_{xx} v_x) \big|_{x=L} - (\kappa u_{xx} v_x) \big|_{x=0} = 0
\end{aligned}
\end{equation}
for all $v$ that satisfy $v(0)=0$.  We can choose successive $v$ such that each term in the above expression can be set equal to zero. Therefore we get the equilibrium equation,
 \begin{eqnarray}
\left( (E u_x - (\kappa u_{xx})_{x}  \right)_{x} + f = 0, 
\end{eqnarray}
additional jump conditions,
 \begin{eqnarray}
[[ E u_x - (\kappa u_{xx})_{x} ]]_{x=l}  \,\,\, \text{and} \,\,\,[[ \kappa u_{xx}  ]]_{x=l},
\end{eqnarray}
and force and natural boundary conditions,
\begin{eqnarray}
(E u_x - (\kappa u_{xx}) - \sigma = 0, \,\,\, (\kappa u_{xx})_{x} (0)=0   \,\,\, \text{and} \,\,\, (\kappa u_{xx})_{x} (L)=0.
\end{eqnarray}

%%%%%%%%%%%%%%%%%%%%%%%%%%%%%%%%%%%%%%%%%%%%%%%
\section{Formulas for the transfer matrix method} \label{app:tm}

The function $\Phi$ in (\ref{eq:itrans0}) is
\begin{eqnarray}
    \Phi = -\begin{pmatrix}
        \lambda^2 e^{x/\lambda}  & \lambda^2 e^{-x/\lambda}  & x & 1 \\
         \lambda e^{x/\lambda}  & -  \lambda e^{-x/\lambda}  & 1 & 0 \\        
        0   & 0  & 1 & 0 \\
        E \lambda^2 e^{x/\lambda} & E \lambda^2 e^{-x/\lambda}  &0 & 0
    \end{pmatrix}.
\end{eqnarray}

The particular solution in (\ref{eq:itrans0}) for $f$ constant is 
\begin{eqnarray}
    \varphi^p = - \frac{1}{E} \left( x^2, x, x, \lambda^2 E \right)^T, 
\end{eqnarray}
while that for $f = e^{ikx}$ is
\begin{eqnarray}
\varphi^p = -\frac{e^{i k x}} {E k^2 + E \lambda^2 k^4}
\left(1, ik, i k + i \lambda^2 k^3, -\lambda^2 k^2 \right)^T.
\end{eqnarray}

%%%%%%%%%%%%%%%%%%%%%%%%%%%%%%%%%%%%%%%%%%%%%%%
\section{Smooth material heterogeneity} \label{app:smooth}

\begin{figure}
\centering
\includegraphics[width=2.5in]{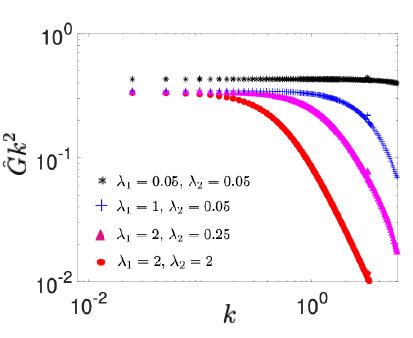}
\caption{ $\hat{G}k^2$ for the smooth material heterogeneity. $E_1=1, E_2=5$.}
\label{fig:smooth}
\end{figure}

We consider the following material heterogeneity,
\begin{equation}
E(x) =  \frac{1}{2} \left(E_1 + E_2 + (E_2-E_1) \sin(2 \pi x) \right), \quad \lambda(x) = \frac{1}{2} \left(\lambda_1 + \lambda_2 + (\lambda_2-\lambda_1) \sin(2 \pi x) \right);
\end{equation}
so $E$ varies smoothly between $E_1$ and $E_2$ while $\lambda$ varies smoothly between $\lambda_1$ and $\lambda_2$.  Figure \ref{fig:smooth} shows the overall response for various choices of material scales.  They have the same qualitative behavior as in Figure \ref{fig:hetper}(c).

%%%%%%%%%%%%%%%%%%%%%%%%%%%%%%%%%%%%%%%%%%%%%%%
\section{Nonlocal elasticiy} \label{app:ke}

We start with the energetic formulation of linear strain gradient elasticity \cite{eringen1972linear}
\begin{equation}
\mathcal{E}(u) = \frac{1}{2} \int_0^L A (E u_x^2 + (K*u_x)^2 - fu )dx + \mathcal{B} \quad \text{where}
\quad (K*v)(x) := \int_0^L K(x-y)v(y) dy.
\end{equation}
The first variation is
\begin{equation}
\delta \mathcal{E} = \left. \frac{d \mathcal{E} ( u+\epsilon v)}{d\epsilon} \right|_{\epsilon=0} = 
 \int_0^L A (E u_x v_x + (K*u_x) (K*v_x) - fv )dx + \mathcal{B}' 
\end{equation}
for a boundary term $\mathcal{B}'$.  Now,
\begin{equation}
\begin{aligned}
 \int_0^L  (K*u_x) (K*v_x) dx  &= \int_0^L \left( (K*u_x)(x) \int_0^L  K(x-y)v_x(y) dy \right) dx \\
 &=  \int_0^L  \int_0^L  \left( (K*u_x)(x) K(x-y)v_x(y) \right) dy  \ dx \\
 &=  \int_0^L  v_x(y) \left(   \int_0^L (K*u_x)(x) K(x-y)  dx \right) dy   \\
&=  \int_0^L  v_x(y) K*(K*u_x) (y) dy .
\ \end{aligned}
\end{equation}
Substituting this back ,
\begin{equation}
\begin{aligned}
\delta \mathcal{E}  &=&  \int_0^L A \left( \left( E u_x + K*(K*u_x) \right)v_x - fv \right) dx + \mathcal{B}' \\
 &=& -\int_0^L A \left( \left( E u_x + K*(K*u_x) \right)_x + f \right) vdx + \mathcal{B}'' 
 \end{aligned}
\end{equation}
for a boundary term $\mathcal{B}''$.  Requiring that $\delta \mathcal{E} = 0$ for all $v$ subject to boundary conditions leads to the governing equation (\ref{eq:nonloc}).

%%%%%%%%%%%%%%%%%%%%%%%%%%%%%%%%%%%%%%%%%%%%%%%
\section{Fractional derivatives and fractional Laplacian} \label{app:fd}

We define the energy in terms of the Gagliardo 2-semi-norm \cite{dinezza_2012}:
\begin{eqnarray}
{\mathcal E}(u) = \int^{L}_{0}A  \left(\int^{L}_{0} \frac{1}{2} M \lambda^{2\alpha} \frac{( u'(x) - u'(y) )^2}{|x-y|^{1+2\alpha}} dy - f(x)u(x) \right)dx + \mathcal{B}
\end{eqnarray}
where we use $'$ to denote derivative.  Taking the first variation,
\begin{equation}
\begin{aligned}
\delta \mathcal{E} &= \left. \frac{d \mathcal{E}(u+\epsilon v)}{d \epsilon} \right|_{\epsilon = 0} \\
&=  \int^{L}_{0} A \left(\int^{L}_{0}  M \lambda^{2\alpha} \left(\frac{ u'(x) - u'(y) }{|x-y|^{1+2\alpha}} \right)(v'(x) - v'(y)) dy - f(x)v(x)  \right) dx  + \bar{\mathcal{B}} \\
&=    \int^{L}_{0} A   \left( 2 M \lambda^{2\alpha}  v' (x) \int^{L}_{0} \left(\frac{ u'(x) - u'(y) }{|x-y|^{1+2\alpha}} \right)  dy - f(x) v(x)  \right) dx  + \bar{\mathcal{B}} \\
&=  -  \int^{L}_{0} A   \left( 2 M \lambda^{2\alpha}  \left( \int^{L}_{0} \left(\frac{ u'(x) - u'(y) }{|x-y|^{1+2\alpha}} \right) dy \right)' + f(x)\right)  v(x)   dx  + \tilde{\mathcal{B}}, 
\end{aligned}
\end{equation}
where we integrate by parts in the final term.  Requiring $\delta \mathcal{E} = 0 $ for all $v$ subject to the boundary condition provides the equilibrium equation in terms for the fractional Laplacian \cite{dinezza_2012},
\begin{equation}
 \left( \frac{2 M \lambda^{2\alpha}}{C} (- \Delta)^\alpha u' \right)' + f = 0 \quad \text{where} \quad (- \Delta)^\alpha u  := C \left(\frac{ u'(x) - u'(y) }{|x-y|^{1+2\alpha}} \right)
\end{equation}
and the constant $C$ depends only on $\alpha$.

\end{document}